\documentclass[%
 reprint,
superscriptaddress,
 amsmath,amssymb,
 aps,
prl,
onecolumn,
11pt
]{revtex4-2}

 \usepackage{color}
 
 \usepackage{stackengine}
\newcommand\xrowht[2][0]{\addstackgap[.5\dimexpr#2\relax]{\vphantom{#1}}}
 
\usepackage{epstopdf}
\usepackage[caption=false]{subfig}

\usepackage{ dsfont }
\usepackage{enumitem}
\usepackage{wrapfig}

\usepackage{graphicx}
\usepackage{amsmath}
\usepackage{amssymb}
\usepackage{mathtools}
\usepackage{multirow}
\usepackage{hyperref}
\usepackage{makecell}
\usepackage{physics}
\usepackage{comment}
\usepackage{tcolorbox}
\usepackage{xcolor}

\usepackage{array}

\newcolumntype{L}[1]{>{\raggedright\let\newline\\\arraybackslash\hspace{0pt}}m{#1}}
\newcolumntype{C}[1]{>{\centering\let\newline\\\arraybackslash\hspace{0pt}}m{#1}}
\newcolumntype{R}[1]{>{\raggedleft\let\newline\\\arraybackslash\hspace{0pt}}m{#1}}

\def\bx{{\bold{x}}}

\def\bG{{\bold{G}}}

\def\bH{{\bold{H}}}

\def\bU{{\bold{U}}}
\def\bV{{\bold{V}}}

\def\m1{{^{-1}}}

\begin{document}


\title{Symmetry Aspects of Chiral Superconductors}

\author{Aline Ramires}

\affiliation{Paul Scherrer Institute, CH-5232 Villigen PSI, Switzerland}

\begin{abstract}
Recent developments in theory, synthesis, and experimental probes of quantum systems have revealed many suitable candidate materials to host chiral superconductivity. Chiral superconductors are a subset of unconventional superconductors which break time-reversal symmetry. Time-reversal symmetry breaking is possible given the order parameter's two-component nature, allowing for a complex relative phase. In this article, we focus on discussing the underlying symmetry aspects that allow for the development of chiral superconductivity. We provide an introductory account of key concepts in group theory and apply these to the classification of order parameters and the generalization of the Landau theory of phase transitions in the context of superconductivity.
\end{abstract}


\maketitle

\section{Introduction}\label{Sec:Introduction}

Superconductivity is one of the most fascinating electronic properties of matter. Materials that are metals at room temperature, transporting electricity at the cost of a finite resistivity, become perfect conductors (with zero resistivity) and perfect diamagnets below a certain temperature. From the discovery of the phenomenon of zero resistivity by Heike Kamerlingh Onnes in 1911 \cite{Onnes1911} and of the Meissner-Ochsenfeld effect (perfect diamagnetism) in 1933 \cite{Meissner1933}, it took physicists a few years to start developing a theoretical understanding of this unexpected phenomena. In 1935, Fritz and Heinz London proposed constitutive relations for a superconductor that, in conjunction with Maxwell's equations, could phenomenologically explain the observed perfect diamagnetism \cite{London1935}. In the later 40s, the Landau theory of phase transitions also found its application in superconductivity, introducing the notion of a superconducting order parameter as a complex number that becomes non-zero below what is known as the superconducting critical temperature \cite{Landau1965}. Only in 1957 Bardeen, Cooper, and Schrieffer (BCS) proposed a microscopic theory with the understanding of superconductivity as a condensate of pairs of electrons that behave as bosons \cite{Bardeen1957}. These pairs of electrons, known as Cooper pairs, are formed by particles with opposite momenta and spin \cite{Cooper1956}.

BCS theory successfully described the phenomenology of the superconducting states in elemental materials and alloys. Surprisingly, since the 70s, multiple families of materials have been discovered with phenomenology that challenges the one initially proposed by the BCS theory: heavy fermions \cite{White2015}, Cu-based \cite{Keimer2015}, and Fe-based \cite{Stuart2011} superconductors,  to name a few. Most of these superconductors were deemed unconventional, and the microscopic theory behind the pairing mechanism remains under active debate in many cases. Even without a consensus on the microscopic pairing mechanism for each material, the phenomenological Landau theory of phase transitions is still a practical framework in the context of unconventional superconductors. In particular, in case the order parameter consists of multiple components, it can give rise to nematic or chiral superconductivity, as we discuss below.

This article focuses on chiral superconductors, a subset of unconventional superconductors. One fundamental aspect of chiral superconductors is the presence of two components that develop with a relative complex phase such that the order parameter breaks time-reversal symmetry, namely, it is not equal to its complex conjugate (for recent reviews on time-reversal symmetry breaking superconductors, see \cite{Wysoki2019, Ghosh2020}). In a chiral superconductor, the phase of the order parameter winds, clockwise or anti-clockwise, around an axis over a loop around the Fermi surface \cite{Kallin2016}. The phase winding is responsible for the time-reversal symmetry breaking (TRSB), the finite angular momentum of the Cooper pair, and nontrivial topology. Intriguingly, TRSB is usually connected with the development of magnetic order, which is often associated with detrimental effects on superconductivity, as it breaks the degeneracy of the electronic states that form the Cooper pair. The experimental observation of superconducting states breaking time-reversal symmetry at the superconducting transition temperature might sound very unlikely initially. However, TRSB at the critical temperature has been observed in multiple superconductors. The most relevant experimental techniques for the direct observation of TRSB are muon spin relaxation ($\mu$SR) \cite{Luke1993, Luke1998, Biswas2021, Biswas2013} (for an introductory survey on the experimental technique, see \cite{Blundell1999}) and polar Kerr effect (PKE) \cite{Xia2006, Schemm2014, Schemm2015} (for an introductory survey on this experimental probe see \cite{Kapitulnik2009}).

In addition to the explicit TRSB, chiral superconductors display other unusual experimental signatures associated with their multi-component character \cite{Sigrist1991}. These include the formation of fractional-quantum vortices, characterized by the distinct winding of the two components in the superconducting order parameter \cite{Jang2011}. Also,  the attenuation of ultrasound in channels with nontrivial symmetry  \cite{Benhabib2021, Ghosh2021}, and the splitting of the superconducting transition temperature and the TRSB temperature under uniaxial strain \cite{Grinenko2021b}. 
Sr$_2$RuO$_4$, a material discovered in the 90's \cite{Maeno1994} displays all these unusual features and 
is the subject of intense theoretical and experimental work \cite{Mackenzie2017, Armitage2020}. 

Chiral superconductors can be topologically nontrivial, exhibiting exotic localized modes at surfaces and lattice defects, surface currents, and thermal Hall effect \cite{Kallin2016, Schnyder2008}. Majorana bound states have been theoretically predicted to be hosted in the vortices of chiral superconductors \cite{Volovik1999, Ivanov2001, Beenakker2013}, and proposed to be used for universal quantum computation \cite{Sarma2015}. For the reader already familiar with second quantization, Kallin and Berlinsky \cite{Kallin2016}, and more recently Sigrist \cite{Sigrist2020}, provide introductory accounts of exceptional phenomena associated with topology in chiral superconductors, including the existence of edge modes, surface currents, fractional-quantum vortices, and thermal Hall conductivity in terms of the Bogoliubov-de Gennes theory. They also introduce the definition of topological invariants, such as the Chern number for systems in two and three dimensions, and highlight its relation to the finite angular momentum of the Cooper pair. Here we are not going to explore topology, so we entrust the reader to these prior works for further reading on this topic.

In this manuscript, we focus on the symmetry aspects of chiral superconductors. We introduce materials and heterostructures recently proposed to host chiral superconductivity highlighting their lattice symmetry in Section \ref{Sec:Materials}. In Section \ref{Sec:Symmetry}, we introduce the minimal set of group theory concepts that allow one to introduce the symmetry classification of order parameters in terms of lattice symmetries. In Section \ref{Sec:PointGroups}, we introduce the notion of point groups and discuss in more detail some of the groups associated with the materials mentioned in Section \ref{Sec:Materials}. We discuss the notion of unconventional superconductivity in Section \ref{Sec:Unconventional}, highlighting the specific properties of nematic and chiral superconductors. In Section \ref{Sec:Landau}, we guide the reader through two examples of the Landau free energy functional construction for multi-component order parameters and discuss the conditions that make chiral superconductivity stable. With this work, we aim to provide an introductory account of the topic of chiral superconductivity that would allow the interested reader to delve into the standard literature in the field with more confidence.

\section{Material Candidates}\label{Sec:Materials}

Unconventional pairing with a chiral character was first mentioned in the context of superfluid $^3$He \cite{Leggertt1975, Wheatley1975}. In fact, superfluid $^3$He has been a forerunner in the context of unconventional phases of matter \cite{VolovikBook}. More generally, unconventional superfluid (or superconducting) states spontaneously break some of the spatial rotation or reflection symmetries present in the normal state, and possibly also time-reversal symmetry. Since the 70's, superfluid $^3$He has been identified to have two distinct superfluid phases in absence of external magnetic fields. At low pressures one finds the B-phase: a superfluid phase consisting of pairs of He atoms with a spin-triplet configuration (distinct from the spin singlet configuration proposed by the BCS theory) and an isotropic gap. At high pressures one finds the A-phase: a chiral superfluid phase, also with spin triplet character, which breaks time-reversal and mirror symmetries. $^3$He continues to be subject of low-temperature research. Recently, controlled disorder and confinement have brought up new superfluid phases \cite{Halperin2018}. The discovery of solid materials potentially hosting chiral superconductivity happened much later. In some cases, there is still a need for better experimental characterization to identify the order parameter as chiral uniquely. 

The discovery of multiple superconducting phases in the heavy fermion material UPt$_3$ \cite{Joynt2002} suggested similarities with the rich phase diagram of superfluid $^3$He and prospects of chiral superconductivity. Broken time-reversal symmetry has been inferred from Josephson interferometry \cite{Strand2009} and PKE \cite{Schemm2014}, even though there are conflicting $\mu$SR results \cite{Luke1993,Dalmas1995}. Neutron scattering experiments could probe the presence of internal degrees of freedom in the superconducting vortices, which corroborates the evidence for a multi-component superconducting order parameter \cite{Avers2020}. Another material in the family of heavy fermions that has been suggested to host chiral superconductivity is URu$_2$Si$_2$ \cite{Mydosh2020}. Chirality in the superconducting state is inferred by direct evidence of TRSB in the superconducting state by $\mu$SR \cite{Schemm2015} and corroborated by analysis of the temperature dependence of the field-angle dependence of specific heat \cite{Kittaka2016}. Most recently, UTe$_2$ has been reported to display multiple superconducting phases as a function of magnetic field and pressure \cite{Ran2019, Ran2021}. Based on unusual asymmetric scanning tunneling spectra, it was suggested that UTe$_2$ hosts a chiral superconducting state \cite{Jiao2020}. This interpretation awaits further experimental confirmation. As we discuss below, a chiral superconducting state is not expected to develop in this material, given the low lattice symmetry.

Analogies to superfluid $^3$He also led to the proposal of chiral superconductivity in Sr$_2$RuO$_4$ \cite{Rice1995}. This suggestion led to the characterization of this material by multiple experimental probes. $\mu$SR \cite{Luke1998} and PKE measurements \cite{Xia2006} give evidence for chiral superconductivity, indicating TRSB below the superconducting critical temperature. Ultrasound attenuation experiments give thermodynamical evidence for a multi-component superconducting order parameter  \cite{Ghosh2021,Benhabib2021}. Furthermore, experiments under uniaxial strain also suggest a two-component order parameter as the superconducting transition temperature and the TRSB temperature are split \cite{Grinenko2021a, Grinenko2021b}. Theoretical calculations that explain the potential origin and nature of this chiral superconducting state were recently proposed \cite{Gingras2019, Suh2020, Clepkens2021}. Despite active theoretical and experimental work on this system for almost three decades, the ultimate symmetry of the order parameter remains to be uniquely identified \cite{Mackenzie2017,Armitage2020}.

SrPtAs, a locally noncentrosymmetric material, is another example of a system displaying  TRSB at the superconducting critical temperature inferred by $\mu$SR \cite{Biswas2013}. Functional renormalization group calculations find that a chiral d-wave superconducting state is favored by the presence of multiple Fermi surfaces in this material \cite{Fischer2014}. Most recently, LaPt$_3$P was also suggested to be a chiral d-wave superconductor based on $\mu$SR experiments and symmetry analysis considering a first-principles band structure \cite{Biswas2021}. Furthermore, the alternate stacking compound 4Hb-TaS$_2$, a natural heterostructure of 1T-TaS$_2$ and 1H-TaS$_2$ layers (the former a Mott insulator and the latter a superconductor) has been proposed to host chiral superconductivity also based on $\mu$SR experiments \cite{Ribak2020}. 

Theoretical works predict chiral superconductivity in many material platforms. The most prominent is graphene, a single carbon sheet, when doped such that the Fermi surface is nested and the Fermi level reaches a van Hove singularity \cite{Pathak2010, Nandkishore2012}. For a review of the different theoretical calculations supporting this proposal, the experimental challenges for its realization, and an extensive report on theoretical proposals of chiral superconductivity in similar materials with hexagonal or trigonal symmetry, see \cite{Black2014}. More recently, chiral superconductivity was also proposed in artificially engineered systems. Twisted double layers of high-critical-temperature copper oxide superconductors were theoretically predicted to host chiral d-wave superconductivity for twist angles in the vicinity of $45^o$ \cite{Can2021}. The same research group has also proposed that a twisted array of proximitized semiconductor nanowires can sustain a chiral superconducting state for twist angles around $90^o$ \cite{Tummuru2021}.

\begin{table}[h]
\begin{center}
    \begin{tabular}{c}
   \includegraphics[width=0.85\linewidth, keepaspectratio]{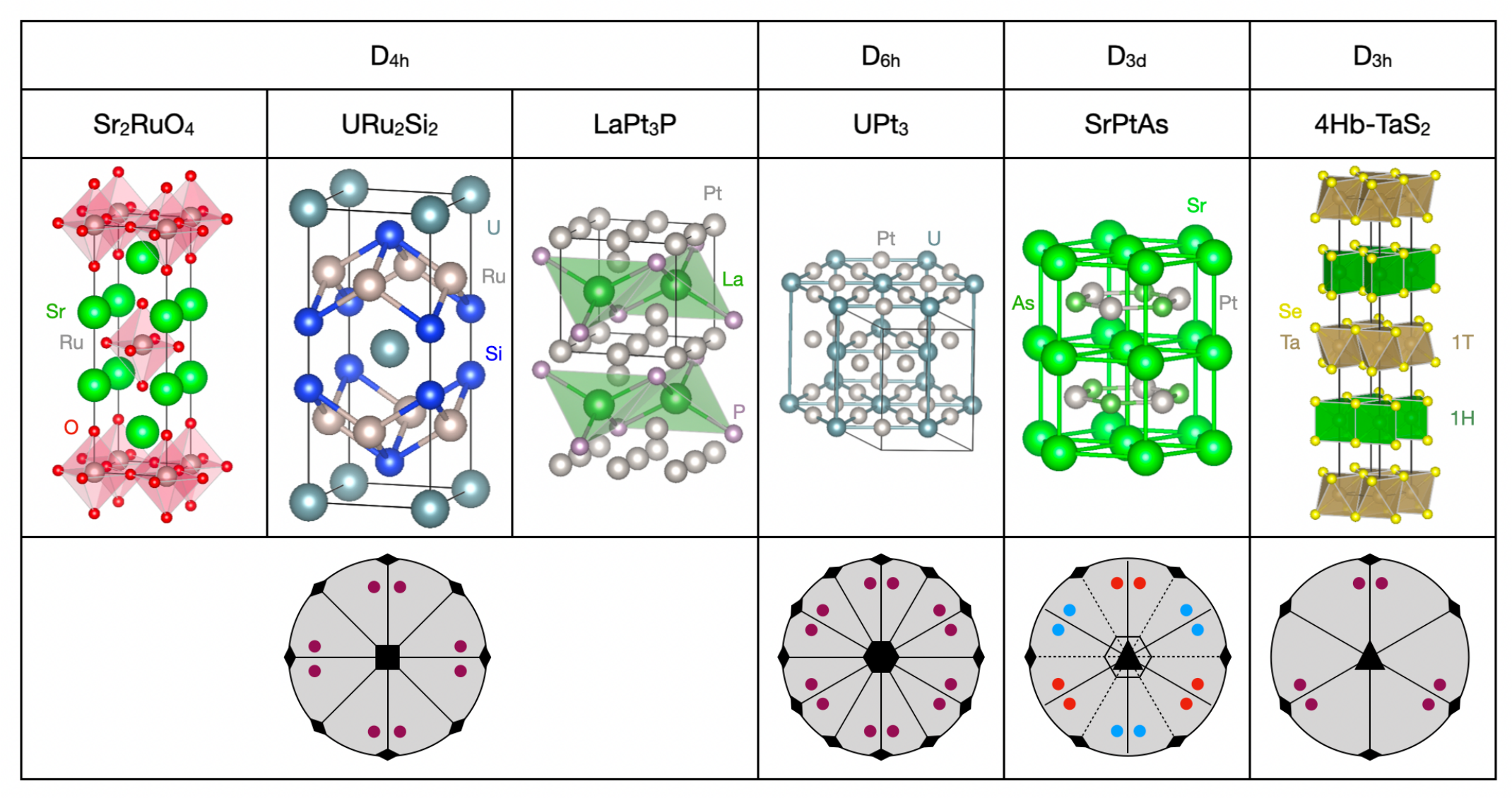}
       \end{tabular}
        \end{center}
\caption{\label{Tab:Materials} Materials proposed to host chiral superconductivity based on direct experimental evidence of TRSB. We highlight the point group symmetry of the crystalline structure in the first line, the chemical formula and the crystalline structure of the respective materials in the second and third lines, and show the  stereographic representation encoding all the symmetry operations of the respective group. On the stereographic projections, a red/blue dot corresponds to a dot above/below the plane of the page, while a purple dot corresponds to pairs of dots above and below the plane of the page. The dots give us information about the presence of horizontal and vertical mirror planes. The number of sides on the polygon at the center of the disk gives the order of the principal rotation axis coming out of the page. The (dotted or full) lines linking the two-fold symmetric rhombuses at the edge of the disks indicate two-fold rotation axes. Full lines correspond to vertical mirror planes. The open hexagon corresponds to a six-fold rotation accompanied by a horizontal mirror reflection. The crystalline structures were generated with the software VESTA \cite{VESTA}.
}
\end{table}

In Table \ref{Tab:Materials}, we highlight the materials proposed to host chiral superconducting states and for which there is direct experimental evidence of TRSB. In the first line, we indicate the point group symmetry; in the second and third lines, we give the chemical formula and the lattice structure of the respective material. In the fourth line, we show the stereographic projection that illustrates all the symmetry elements of the respective point group in a concise manner. 

The point group labels give us the following information: all are dihedral groups (corresponding to the letter $D$), which means that these are the symmetry groups of regular polygons with $n$-sides (corresponding to the numerical subindex). Regular polygons of $n$ sides have a principal $n$-fold rotation axis plus $n$ two-fold rotation axes perpendicular to the principal axis. The last subindex tells us about additional symmetries: a subindex $h$ corresponds to a (horizontal) mirror plane perpendicular to the principal rotation axis, and a subindex $d$ corresponds to the presence of mirror planes containing the $n$-fold axes.

In the next section, we will introduce the formal notion of a group and the minimal set of concepts that allow us to classify order parameters in terms of irreducible representations of a group. In Section \ref{Sec:PointGroups}, we will go back to the specific properties of the mentioned groups in more detail.

\section{Symmetry aspects}\label{Sec:Symmetry}

A good starting point to discuss symmetry aspects of superconducting states is remembering that superconductivity develops in real materials with an associated \emph{crystalline lattice} structure with specific symmetry properties. The symmetries of a crystalline lattice are characterized by the corresponding \emph{space group}, the group of transformations that leave the lattice invariant. These transformations can include translations, rotations, reflections, and their combinations. If we focus on transformations that leave one point of space invariant, we have what is called a \emph{point group}, which includes rotations and reflections (but excludes translations). Before introducing the specifics of point groups (here, we do not discuss space groups), we will learn some basic ideas about group theory more abstractly. Group theory has many applications in physics, not only in condensed matter physics but also in high energy physics. Here we will introduce the minimum set of concepts necessary to understand the classification of superconducting order parameters. Classical textbooks such as  \cite{Georgi, Hamermesh, Bradley} provide a complete introduction to group theory.

For didactic purposes, here we start with the example of the symmetry group of the equilateral triangle to introduce the main concepts. We hope that the interested reader can follow the same steps to construct and understand the  tables for the symmetry groups associated with the materials mentioned in Section \ref{Sec:Materials}. 

\subsection{A minimal introduction to group theory}

We start with the most abstract definition of a \emph{group}:

\begin{tcolorbox}[width=0.95\textwidth,center,colframe=white,colback=lightgray]    
\textbf{Definition:} A \emph{group} $\bG$ is a set of elements together with a composition law (.), also referred to as \emph{product} or \emph{multiplication law}, such that:

\begin{enumerate}

\item The product of any two elements is a member of the group: 

if $A$ and $B \in \bG$, then $C = A.B\in \bG$;

\item The product is associative: 

$A.(B.C) = (A.B).C$ for all $A, B, C \in \bG$;

\item There exists a unique identity element $I$: 

$I.A=A.I=A$ for all $A \in \bG$;

\item Every element has a unique inverse: 

given $A\in\bG$, there exists an element $A^{-1}$ such that $A.A^{-1}= A^{-1}.A =I$.

\end{enumerate}
\end{tcolorbox}  

This definition is somewhat abstract, so let us consider one example of such a mathematical structure:
\\

\textbf{Example A:} The integer numbers $\mathds{Z} = (...-2,-1,0,1,2,...)$ with the operation of addition (+) is called the \emph{additive group of the integers}. Note that it satisfies all requirements above:

\begin{itemize}[label={}]

\item (1A) The composition law (here addition) of any two elements is a member of the group: 

$1+1=2$, $2+7=9$, $(-5)+3=(-2)$, $(-1)+(-3)=(-4)$,...

\item (2A) The composition is associative: 

$1+5+(-3) = (1+5)+(-3) = 6+(-3) = 3$

$1+5+(-3) = 1+[5+(-3)] = 1+2 = 3$

\item (3A) There exists a unique identity element $I=0$: 

$1+0=1$, $3+0=3$, $(-5)+0=(-5)$,...

\item (4A) Every element has a unique inverse (the element with the opposite sign): 

$1 + (-1)=0$, $(-3)+3 = 0$,...

\end{itemize}

This example is helpful to note that groups are more general mathematical structures, not necessarily associated with symmetry operations, as we will be focusing on below.
\\

\begin{figure}[h]
\begin{center}
\includegraphics[width=0.85\linewidth, keepaspectratio]{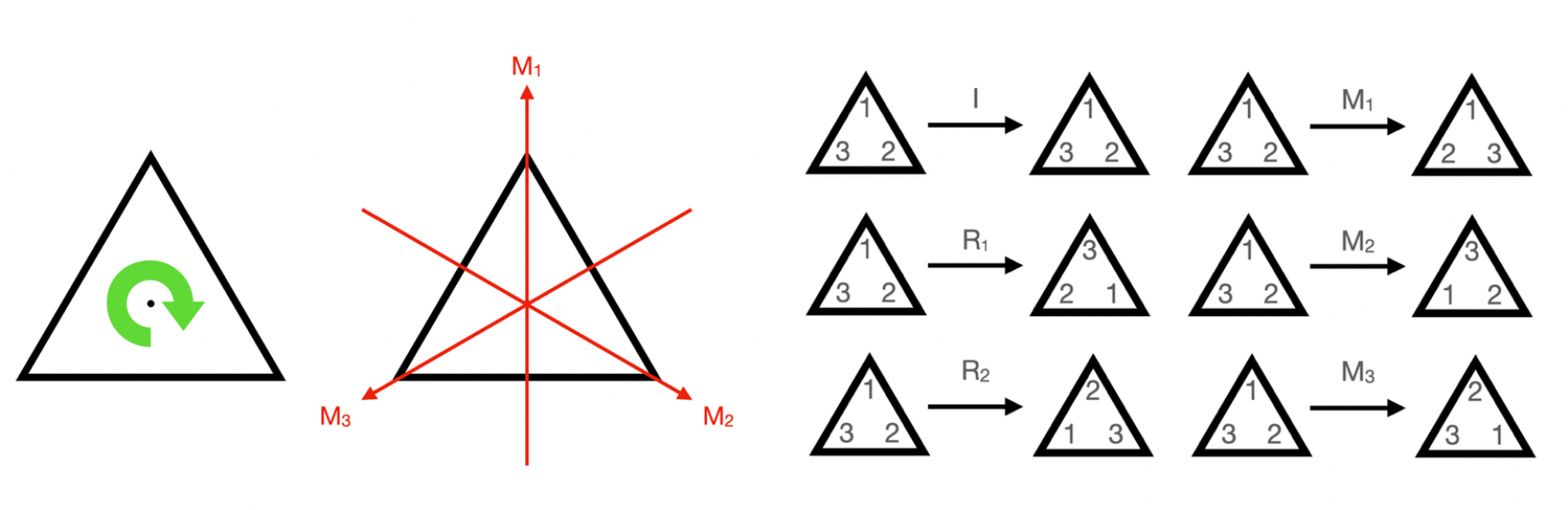}
\caption{\label{Fig:Triangle} Symmetry operations on the equilateral triangle. Left: Clockwise rotation around the center of the triangle. Middle: Mirror planes. Right: explicit transformation rules can be followed by labeling the triangle edges with numbers 1, 2, and 3. }
\end{center}
\end{figure}

\textbf{Example B:} The symmetry operations of the equilateral triangle form a group. The group is composed of the following operations: identity, $I$, clockwise rotations by $120^o$ ($R_1$) and $240^o$ ($R_2$) around the axis passing through the center of the triangle (going into the page), and reflections at three different mirror planes which pass through the center and one of the triangle's edges: $M_i$ with $i=1,2,3$, as indicated in Figure \ref{Fig:Triangle}. This group is also known as $C_{3v}$, as it has one three-fold symmetric rotation axis and three vertical mirror planes.

Note that the set formed by these operations, $\{I,R_1,R_2,M_1,M_2,M_3\}$,  satisfy all the requirements for a group:

\begin{itemize}[label={}]

\item (1B) The ``product"  (here the composition) of any two elements is a member of the group (as a convention, the rightmost operation is the one to be applied first). Some examples:

Composition of two rotations: $R_1.R_1 = R_2$, $R_1.R_2 = I$, $R_2.R_1 = I$, $R_2.R_2=R_1$. 

Composition of two mirror operations: $M_i.M_i = I$ for $i=1,2,3$, $M_1.M_2 = R_1$, $M_2.M_1 = R_2$, and so on. 

Compositions of rotations and mirror operations: $R_1.M_1 = M_3$, $M_1.R_1 = M_2$, and so on. 

There is a total of $6^2=36$ pairs of operations to be verified. The reader can check that all combinations of two operations result in one of the six operations in the set, as summarized in the \emph{multiplication table} shown in Table \ref{Tab:MTExB}.

\item (2B) The product is associative: 

$M_1.R_1.M_3 = (M_1.R_1).M_3 = M_2.M_3 = R_1$

$M_1.R_1.M_2 = M_1.(R_1.M_3) = M_1.M_2= R_1$

This is one example of the associative property. There is a total of $6^3=216$ triples to be verified. The reader can check that this applies to any choice of three telements in the group.

\item (3B) There exists a unique identity element $I$: here is the operation equivalent to ``doing nothing" on the triangle.

$I. R_i = R_i . I = R_i$, for $i=1,2$

$I. M_i =M_i . I = M_i$, for $i=1,2,3$

\item (4B) Every element has a unique inverse: 

$R_1.R_2 = R_2.R_1= I$

$M_i.M_i = I$ for $i=1,2,3$

This property is manifest in Table \ref{Tab:MTExB} by the fact that the element $I$ appears only once in each row and column.

\end{itemize}

\begin{table}[h]
\begin{center}
    \begin{tabular}{| c | c | c | c | c | c | c |}
    \cline{2-7}
\multicolumn{1}{c|}{} &   $\bold{I}$ & $\bold{R_1}$ & $\bold{R_2}$ & $\bold{M_1}$ & $\bold{M_2}$ & $\bold{M_3}$ \\ \hline
 $\bold{I} $&$I$ & $R_1$ & $R_2$ & $M_1$ & $M_2$ & $M_3$ \\ \hline
$\bold{R_1}$& $R_1$ & $R_2$ & $I$ & $M_3$ & $M_1$ & $M_2$ \\ \hline
$\bold{R_2}$& $R_2$ & $E$ & $R_1$ & $M_2$ & $M_3$ & $M_1$ \\ \hline
$\bold{M_1}$ & $M_1$ & $M_2$ & $M_3$ & $I$ & $R_2$ & $R_1$ \\ \hline
$\bold{M_2}$ & $M_2$ & $M_3$ & $M_1$ & $R_1$ & $I$ & $R_2$ \\ \hline
$\bold{M_3}$ & $M_3$ & $M_1$ & $M_2$ & $R_2$ & $R_1$ & $I$ \\ \hline
    \end{tabular}
        \end{center}
    \caption{  \label{Tab:MTExB} Multiplication table for the group of symmetries of the equilateral triangle. The entries in the first line give the operation applied first,  and the entries in the first column the operation applied second. The corresponding entry gives us the resulting operation. For example: applying first $M_2$ and then $M_1$ we find $R_1$, or $M_1.M_2 = R_2$. }
\end{table}

For our discussion, the definitions below are also going to be useful:

\begin{tcolorbox}[width=0.95\textwidth,center,colframe=white,colback=lightgray]    

 \textbf{Order of a group:} The number of elements in a group;
 \\

\textbf{Order of an element $A$:} The least positive integer $s$ such that $A^s=I$;
\\

\textbf{Generators:} A subset of elements in the group that generates the entire group by taking products between them. More precisely, the elements $P_1,P_2,...,P_m$ of a group $\bG$ (with order larger or equal to $m$) are called generators if every element in $\bG$ can be expressed as a finite product of powers of these elements. The choice of generators is not unique;
\\

\textbf{Subgroup:} A subset $\bH$ of a group $\bG$ that is itself a group under the same composition law as in $\bG$;
\\

\textbf{Conjugate Elements:} Two elements $G_1$ and $G_2$ are said to be conjugate if there exists an element $G$ in $\bG$ such that $G_1 = G G_2 G^{-1}$;
\\

 \textbf{Conjugacy classes:} The elements of a group are split into conjugacy classes $C_1, C_2,C_3,...$ such that the following properties hold:

\begin{itemize}[label={}]

\item (i) Every element of $\bG$ is in some class and no element of $\bG$ is in more than one class such that $\bG = C_1+C_2+C_3+...$;

\item (ii) All elements in a given class are mutually conjugate and consequently have the same order;

\item (iii) An element that commutes with all other elements of the group is on a class by itself;

\item (iv) The number of elements in a class is a divisor of the order of the group;

\end{itemize}

\end{tcolorbox}

Let us now discuss how these definitions apply to Example B. The order of the group is six. Rotations are operations of order three, and mirror reflections are of order two. The multiplication table, Table \ref{Tab:MTExB}, organizes the result of the multiplication of all pairs of elements in the group. We can choose two operations $R_1$ and $M_1$, for example, and show that the entire group can be generated from products of these: $I = R_1.R_1.R_1 = M_1.M_1$, $R_2 = R_1.R_1$, $M_2 = M_1.R_1$, and $M_3 = R_1.M_1$.  Note that the choice of generators is not unique. We could have as well chosen $R_2$ and $M_3$ and also generate the entire group: $I = R_2.R_2.R_2 = M_3.M_3$, $R_1 = R_2.R_2$, $M_1 = R_2.M_3$, and $M_2 = M_3.R_2$.

To determine the conjugacy classes, we start with the identity element $I$ forming a class by itself. For any element $G$ in the group $G .I .G^{-1} = G.G^{-1}.I = I$, such that the identity element is not conjugate to any other element in the group and forms a conjugacy class by itself, $C_1=\{I\}$. Considering rotations, we see that a composition of two rotations is always a rotation and that  $M_i. R_1 . M_i^{-1} = M_i. R_1 M_i = R_2$ (for all $i=1,2,3$), so the elements $R_1$ and $R_2$ are conjugate and form a second conjugacy class $C_2 = \{R_1,R_2\}$. Considering  mirror operations, we find that $R_1.M_a.R_1^{-1} = R_1.M_a.R_2 = M_b$, true for $(a,b)=\{(1,2),(2,3),(3,1)\}$, such that the mirror elements form a third conjugacy class among themselves $C_3 = \{M_1,M_2,M_3\}$. Note that these conjugacy classes follow all the properties (i)-(iv) enumerated above. Note also that conjugacy classes join operations of the same ``type": rotations are conjugate to rotations, and mirror reflections are conjugate to mirror reflections.

\subsection{Group representations}

So far, we have explored group theory in a rather abstract way. Example B comes closer to what we are interested in, with the group of symmetry operations of a given object in real space. We can now think more concretely in real space using cartesian coordinates and matrices acting on the coordinates to represent symmetry operations. The symmetry operations of the equilateral triangle acting on a point with coordinates $P=(x,y,z)$ are:
\begin{eqnarray}\label{Eq:3DRep}
D_{3D}(I) = \begin{pmatrix}
1 & 0 & 0 \\
0 & 1 & 0 \\
0 & 0 & 1
\end{pmatrix},&&
\hspace{0.5cm}
D_{3D}(R_1) = \begin{pmatrix}
-1/2 & +\sqrt{3}/2 & 0 \\
-\sqrt{3}/2 & -1/2 & 0 \\
0 & 0 & 1
\end{pmatrix},
\\ \nonumber
D_{3D}(M_1)& =& \begin{pmatrix}
-1 & 0& 0 \\
 0 & 1 & 0 \\
0 & 0 & 1
\end{pmatrix}.
\end{eqnarray}
Here $D_\ell(X)$ corresponds to a matrix representation with label $\ell$ of operation $X$. The label $3D$ corresponds to three-dimensional space.

\begin{tcolorbox}[width=0.95\textwidth,center,colframe=white,colback=lightgray]    

\textbf{Definition:} A \emph{representation} of a group $\bG$ is a mapping $D_\ell$ of the elements of $\bG$ onto a set of linear operators (matrices) with the following properties:

(I) $D_\ell(I) = 1$, where $1$ is the identity operator on the space in which the linear operator acts;

(II) $D_\ell(G_1)D_\ell(G_2) = D_\ell(G_1G_2)$, meaning that the group multiplication law is preserved under the mapping.

\end{tcolorbox}

The reader can explicitly check that the compositions of the matrices in Eq. \ref{Eq:3DRep} emulate the multiplication table displayed in Table \ref{Tab:MTExB}. This construction is associated with the concept of  \emph{group representations}, for which the notion of characters is handy:

\begin{tcolorbox}[width=0.95\textwidth,center,colframe=white,colback=lightgray]    

\textbf{Character:} The characters of a group representation $D_\ell$ are the traces of the respective linear operators (matrices) $\chi_{\ell}(G_i) = \text{Tr}[D_{\ell}(G_i)]$. The trace of a matrix is the sum of its diagonal elements.

\end{tcolorbox}

From the definition above, it follows that if $G_1$ and $G_2$ are in the same conjugacy class, they have the same trace. If $G.G_1.G^{-1} = G_2$, then  $\chi(G_2) = \chi(G.G_1.G^{-1}) = \chi (G^{-1}.G.G_1) = \chi(G_1)$, using the cyclic property of the trace. Note that, by the same property, any similarity transformation (or change of basis) does not change the characters and therefore preserves the conjugacy class structure.

Going back to Example B, the characters of the three-dimensional matrix representation proposed above for the symmetry group of the equilateral triangle are the following: $\chi_{3D}(E) = 3$, $\chi_{3D}(R_1,R_2) = 0$, $\chi_{3D}(M_1,M_2,M_3)$ = 1. Note that we already used that rotations (mirror reflections) belong to the same conjugacy class and therefore have the same character.

\subsubsection{Irreducible representations}

We now introduce the key concept of \emph{reducible} and \emph{irreducible representation} (irrep). For that, we start with the definition of invariant subspace.

\begin{tcolorbox}[width=0.95\textwidth,center,colframe=white,colback=lightgray]    
\textbf{Definition:} Let $D_\ell$ be a representation of a group $\bG$ acting on a vector space $\bV$. $\bU$ is said to be an \emph{invariant subspace} of $\bV$ under $D_\ell$ if: 

($\alpha$)  $\bU$ is a vector subspace of $\bV$;

($\beta$) $D_\ell(G) \bx \in \bU$ for all $G$ in $\bG$ and $\bx \in \bU$.
\end{tcolorbox}

If $\bV$ has no proper invariant subspace under $D_\ell(G)$ (except the zero-vector and $\bV$ itself), then $D_\ell(G)$ is said to be an \emph{irreducible representation}. If there exists a proper invariant subspace under $D_\ell(G)$ then $D_\ell$ is said to be \emph{reducible}. Suppose $\bV$ can be split up into the direct sum of subspaces, each of which is invariant under $D_\ell(G)$ and each of which is the carrier space for an irreducible representation of $\bG$. In that case, $D_\ell$ is said to be \emph{completely reducible}.

To understand better the definitions above, we can go back to the example of the symmetry group of the equilateral triangle with the explicit matrix representation constructed for the three-dimensional space given by Eq. \ref{Eq:3DRep}. We see from these matrices that the rotations mix the x- and y-components, but the z-component never mixes with the other components by any symmetry operation. This segmentation allows us to divide the space in (x,y) and (z) as two invariant subspaces and write smaller operations representations for each. In this context, we say that the representation provided in Eq. \ref{Eq:3DRep} is reducible.

Acting on the two-dimensional space (x,y), we have the first irreducible representation $D_E$ (we discuss the meaning of the labels of the irreducible representations below):
\begin{eqnarray}
D_E(I)= \begin{pmatrix}
1 & 0  \\
0 & 1 
\end{pmatrix},
\hspace{1cm}
D_E(R_1) = \begin{pmatrix}
-1/2 & +\sqrt{3}/2  \\
-\sqrt{3}/2 & -1/2 
\end{pmatrix},
\hspace{1cm}
D_E(M_1) = \begin{pmatrix}
-1 & 0  \\
 0 & 1
\end{pmatrix}.
\end{eqnarray}

The characters in this case are $\chi_{E}(I) = 2$, $\chi_{E}(R_1,R_2) = -1$, $\chi_{E}(M_1,M_2,M_3)$ = 0.

Acting on the one-dimensional space (z), we have the second irreducible representation:
\begin{eqnarray}\label{Eq:Trivial}
D_{A_1}(I) = 1,
\hspace{1cm}
D_{A_1}(R_1) = 1,
\hspace{1cm}
D_{A_1}(M_1) = 1.
\end{eqnarray}

The characters in this case are $\chi_{A_1}(I) = 1$, $\chi_{A_1}(R_1,R_2) = 1$, $\chi_{A_1}(M_1,M_2,M_3)$ = 1. Equation \ref{Eq:Trivial} displays the \emph{trivial irrep}: all elements are mapped into the one dimensional identity matrix.

Are these all the possible irreducible representations? In order to check if we have identified all irreducible representations, we can verify the properties highlighted below (proofs of these properties can be found in \cite{Georgi, Hamermesh, Bradley}).

\begin{tcolorbox}[width=0.95\textwidth,center,colframe=white,colback=lightgray]    
Properties of irreducible representations:

\begin{itemize}[label={}]

\item (a) The number of irreducible representations, $r$, is equal to the number of conjugacy classes;

\item (b) The order of the group $\bG$, $|\bG|$, is equal to the sum of the squares of the dimensions of the irreducible representations $d_i$, $|\bG|=\sum_{i=1}^r d_i^2$;

\item (c) The characters are orthonormal: $\sum_{i=1}^r n_i \chi_D^{*}(G_i)\chi_{D'}(G_i) = |\bG| \delta^{DD'}$, where $n_i$ is the number of elements in the conjugacy class represented by $G_i$. Here $^{*}$ corresponds to complex conjugation.

\end{itemize}
\end{tcolorbox}

For the example of the symmetry transformations of the equilateral triangle, we have found two irreps. We have concluded above that there are three conjugacy classes $C_{i=1,2,3}$, so given property (a), we are missing one irrep. From property (b), we can infer the dimension of the irrep: $6=1^2+2^2 + d^2$, so the dimension of the third irrep must be $d=1$. Using property (c), we can find the characters of this new irrep (in this case, it corresponds to the matrixes themselves, as these are one-dimensional). The orthogonality condition with the trivial ($A_1$) and with the two-dimensional ($E$) irreps identified above give us:
\begin{eqnarray}
(1).1.1 + (2).1.\chi_{A_2}(R_i) + (3).1.\chi_{A_2}(M_i) = 0
\\
(1).2.1 + (2).(-1).\chi_{A_2}(R_i) + (3).0.\chi_{A_2}(M_i) = 0.
\end{eqnarray} 

Solving these two equations we find: $\chi_{A_2}(R_i)=1$ and $\chi_{A_2}(M_i)=-1$. The third irreducible representation then reads:
\begin{eqnarray}
D_{A_2}(I) = 1,
\hspace{1cm}
D_{A_2}(R_1)= 1,
\hspace{1cm}
D_{A_2}(M_1) = -1.
\end{eqnarray}

The \emph{character table}, displayed in Table \ref{Tab:CTC3v},  organizes all the information about the irreducible representations and their characters. Note that with properties (a) - (c), we could have constructed the character table without going through the three-dimensional representation based on euclidean coordinates $(x,y,z)$. We know from the abstract structure of the group that there are six elements divided into three conjugacy classes. This tells us that there are 3 irreducible representations $D_{i=a,b,c}$. The dimensions of the irreducible representations, $d_{i=a,b,c}$ should follow the identity $6 = d_a^2 + d_b^2 + d_c^2$. Since each dimension is a positive integer, the only option is to have $d_a=1$, $d_b=1$ and $d_c =2$. One of the one dimensional irreps is always the trivial representation, with $\chi_{a} (G_j) = 1$ for all elements $G_j$ of $\bG$ (this irrep corresponds to the irrep labelled as $A_1$ in Table \ref{Tab:CTC3v}). Now we can use the orthogonality of the characters to determine the other irreps. For the second one-dimensional irrep we have $\chi_{b}(I)=1$. The orthogonality condition with the first irrep reads: $(1)1.1 + (2).1.\chi_{b}(R_i) + (3).1.\chi_{b}(M_i) = 0$, and the normalization condition reads $(1)1^2 + (2).[\chi_{b}(R_i)]^2 + (3).[\chi_{b}(M_i)]^2 = 6$, what leads to $\chi_{b}(R_i) = 1$ and $\chi_{b}(M_i) = -1$, so we have completed the second line of the table, labeled as $A_2$. Moving now to the two-dimensional irrep, we know that $\chi_c(I) = 2$. We can again use the orthogonality conditions. The orthogonality with respect to the first irrep gives: $(1)1.2 + (2).1.\chi_{c}(R_i) + (3).1.\chi_{c}(M_i) = 0$. The orthogonality with respect to the second irrep gives: $(1)1.2 + (2).1.\chi_{c}(R_i) + (3).(-1).\chi_{c}(M_i) = 0$. So we conclude that $\chi_{c}(M_i) = 0$ and $\chi_{c}(R_i) = -1$, completing the third row of character table, labeled as $E$. In the discussion based on the three spatial coordinates above, we then identify $a=1$, $b=3$, and $c=2$. 

In Table \ref{Tab:CTC3v}, we label the irreps according to the standard notation. Labels $A$ and $B$ indicate one-dimensional irreps with character $+1$ or $-1$ for the rotations along the principal axis. $E$ indicates two-dimensional irreps, and $T$ indicates three-dimensional irreps. Since here there are two one-dimensional irreps with character $+1$ for the rotations along the principal axis, we distinguish them by the indexes $1$ and $2$. Here $A_1$ labels the trivial irrep. For groups with inversion symmetry, irreps are further labeled by their parity with subindex $g$ or $u$ for even or odd irreps (from \emph{gerade} or \emph{ungerade} in German), as we are going to discuss below.

\begin{table}[h]
\begin{center}
    \begin{tabular}{c}
   \includegraphics[scale=0.4, keepaspectratio]{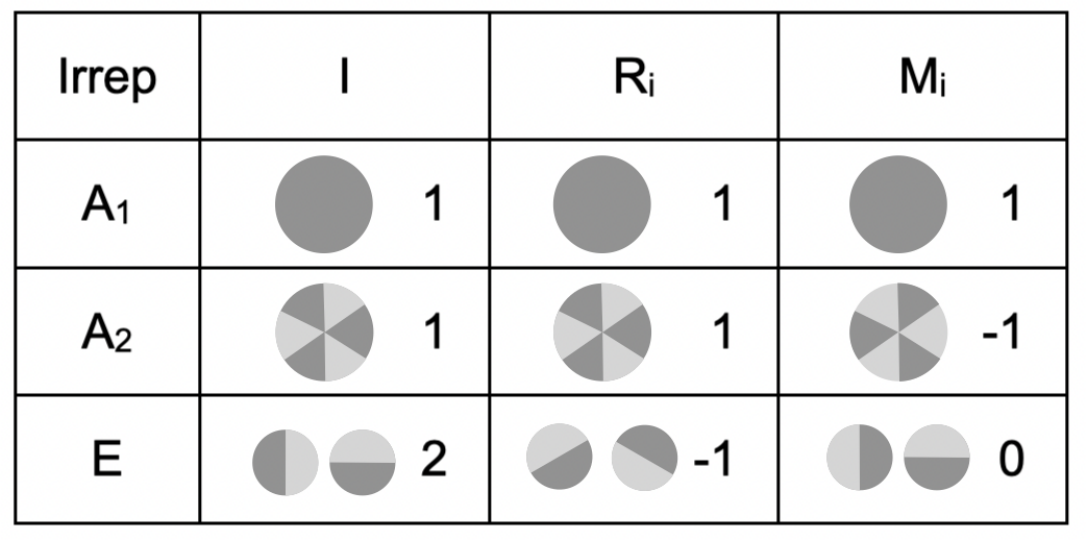}
       \end{tabular}
        \end{center}
\caption{\label{Tab:CTC3v} Illustrated character table of the point group $C_{3v}$, the group of symmetries of the equilateral triangle. In this table, we display images accompanying the numerical entries found in the standard tables. The dark/bright gray colors indicate regions associated with the +1/-1 value. The circles are in the $xy$-plane.}
\end{table}

Irreps can be understood as the different ``classes" of objects (or functions), classified according to how they transform under all operations in the group. Note that the existence of irreducible representations with a dimension larger than one (in the example of the group $C_{3v}$ the irrep labeled as $E$) stems from the fact that two coordinates mix under some of the transformations (here a 3-fold rotation along the z-axis necessarily mixes the x- and y-coordinates). This is a general aspect that we encounter in other point groups.

\section{Crystallographic point groups}\label{Sec:PointGroups}

As we have introduced earlier, a point group is a group of symmetry operators that leave a point $P$ fixed, preserving all distances and angles in Euclidean space. The symmetry operations can include rotations about an axis passing through $P$ or reflections in planes containing $P$. So far, we have focused on the example of the equilateral triangle, with the center of the triangle being the fixed point. 

Now we will focus on crystallographic point groups, namely, groups associated with possible crystal structures of real materials. For these groups, besides the requirement of leaving a point fixed, the symmetry operations also need to be consistent with translational symmetry (even though translations are not in the group themselves). This requirement allows axes of rotation that are only 2-, 4-, or 6-fold. It is possible to demonstrate that there are only ten crystallographic point groups in two dimensions and thirty-two crystallographic point groups in three dimensions \cite{Bradley}.

From Table \ref{Tab:Materials}, we see that the materials that have been suggested to support chiral superconductivity have point groups $D_{4h}$,  $D_{6h}$, $D_{3h}$, or $D_{3d}$. Below we introduce in detail groups $D_{4h}$,  $D_{6h}$ and display the illustrated character tables for $D_4$ and $D_6$, the corresponding subgroups eliminating inversion symmetry. We also comment on how information about groups $D_{3h}$, or $D_{3d}$, can be inferred from the character table of $D_{6h}$.

The point group $D_{4h}$ is composed of the following operations: $I$, the identity; rotations by $\pm \pi/2$ around the z-axis, denoted as $2C_4(z)$ [corresponding to $C_4(z)$ and $C_4^{-1}(z)$]; rotation by $\pi$ around the z-axis, $C_2(z)$; rotation by $\pi$ around the x- and y-axes, denoted as $2C_2(x)$ [corresponding to $C_2(x)$ and $C_2(y)$]; rotation by $\pi$ around the diagonal and antidiagonal, denoted as $2C_2(d)$  [corresponding to $C_2(d)$ and $C_2(\bar{d})$]; inversion, $i$; screw rotations consisting of $\pi/2$ rotations along the z-axis followed by a xy-mirror plane reflection, denoted as $2S_4$ (corresponding to $S_4$ and $S_4^{-1}$); $xy$-mirror reflection $\sigma_h$, two vertical mirror reflections passing through the $xz$- and $yz$-planes, denoted as $2\sigma_v$ (corresponding to $\sigma_{xz}$ and $\sigma_{yz}$); and two vertical mirror planes along the diagonals, denoted by $2\sigma_d$ (corresponding to $\sigma_{d}$ and $\sigma_{\bar{d}}$). These sixteen symmetry operations form ten conjugacy classes. Therefore the group $D_{4h}$ has ten irreducible representations. Given the presence of inversion symmetry, these can be separated into five even, subindex $g$, and five odd, subindex $u$ irreps. 

To illustrate the main features of this approach, we choose to restrict our discussion to the simpler case of $D_4$, the subgroup of $D_{4h}$ if we eliminate inversion symmetry $i$. $D_4$ has only half the symmetry operations of $D_{4h}$. Conversely, $D_{4h}$ can be considered the group $D_4$ extended by the operation of inversion symmetry. There are twice as many operations as in $D_4$: the original operations plus all the operations composed with inversion. There are now only eight symmetry operations organized in five conjugacy classes. Therefore there are five irreducible representations (not labeled as $u$ or $g$, as inversion symmetry is not present). The irreducible representations of $D_{4h}$ can be thought of as the same irreducible representations of $D_4$ but split into even ($g$) and odd ($u$). Table \ref{Tab:CTD4} shows the character table of $D_4$ with its five irreducible representations. The corresponding table for $D_{4h}$ would be four times larger, but the information conveyed would be the same up to identifying the irrep as even or odd. The explicit mapping of irreps follows: $A_{1g/u} \rightarrow A_1$, $A_{2g/u} \rightarrow A_2$, $B_{1g/u} \rightarrow B_1$, $B_{2g/u} \rightarrow B_2$, and $E_{g/u}\rightarrow E$. The complete character table for $D_{4h}$ (and all other crystallographic point groups) can be found in books such as \cite{Bradley} or on websites such as \cite{Katzers}. Note that most of the character tables include basis functions associated with different irreps. These as functions of the coordinates $(x,y,z)$ which transform according to the characters, in the same way as the objects displayed on the tables. For example, we have the following lowest power basis functions for the $D_4$ point group: $f_{A_1} = cte$, $f_{A_2} = z$, $f_{B_1} = x^2-y^2$, $f_{B_2} = xy$, and $f_{E} = \{x,y\}$. The corresponding lowest power basis functions associated with the irreps of the $D_{4h}$ point group are: $f_{A_{1g}} = cte$, $f_{A_{2g}} = xy (x^2-y^2)$, $f_{B_{1g}} = x^2-y^2$, $f_{B_{2g}} = xy$, and $f_{E_g} = \{xz,yz\}$, for the even sector, and $f_{A_{1u}} = xyz (x^2-y^2)$, $f_{A_{2u}} = z$, $f_{B_{1u}} = xyz$, $f_{B_{2u}} = z(x^2-y^2)$, and $f_{E_u} = \{x,y\}$ for the odd sector.

\begin{table}[h]
\begin{center}
    \begin{tabular}{c}
\includegraphics[scale=0.4, keepaspectratio]{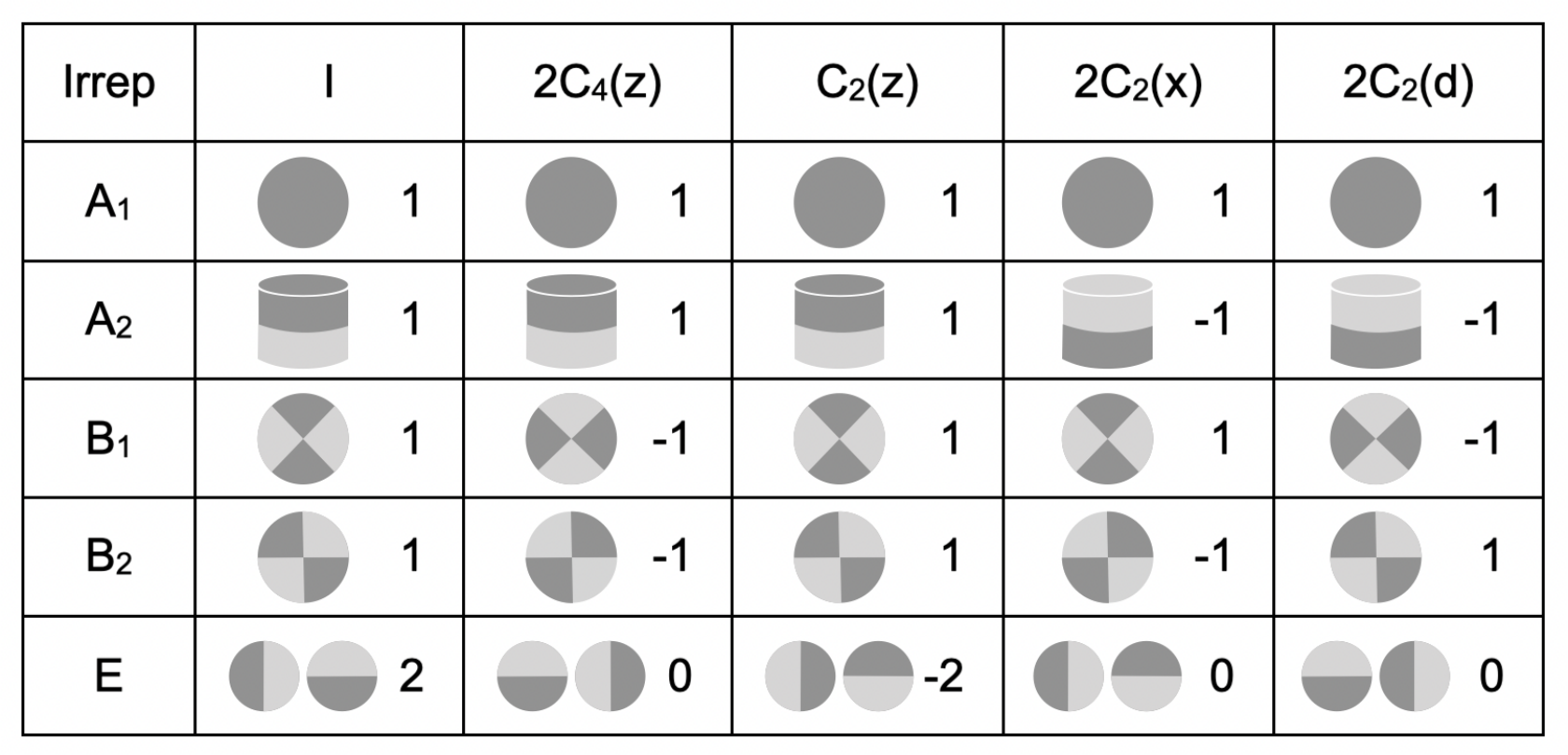}
       \end{tabular}
        \end{center}
\caption{\label{Tab:CTD4} Illustrated character table of the group $D_4$ with images accompanying the numerical entries. The dark/bright gray colors indicate regions with a +1/-1 value. The circles are in the $xy$-plane and the cylinder  axis is along the $z$-axis.}
\end{table}

The point group $D_{6h}$ is composed of the following operations: $I$, the identity; rotations by $\pm \pi/3$ around the z-axis, denoted as $2C_6(z)$ [corresponding to $C_6(z)$ and $C_6^{-1}(z)$]; rotations by $2\pi/3$ around the z-axis, denoted as $2C_3(z)$ [corresponding to $C_3(z)$ and $C_3^{-1}(z)$]; rotation by $\pi$ around the z-axis, $C_2(z)$; rotations by $\pi$ around the x-axis and the other three corresponding axis related by $C_3(z)$ rotations, denoted as $3C_2(x)$; rotations by $\pi$ around the y-axis and the other three corresponding axis related by $C_3(z)$ rotations, denoted as $3C_2(y)$; inversion, $i$; screw rotations consisting of $\pi/3$ rotations along the z-axis followed by a xy-mirror plane reflection, denoted by $2S_6$ (corresponding to $S_6$ and $S_6^{-1}$);  $xy$-mirror reflection $\sigma_h$; three vertical mirror reflections passing through the $xz$- and and other planes related by $C_3(z)$ rotations, denoted as $3\sigma_v$; and three vertical mirror planes along the $yz$-plane, denoted by $3\sigma_d$. These twenty-four symmetry operations form 12 conjugacy classes. Therefore the group $D_{6h}$ has 12 irreducible representations. Given the inversion symmetry, these can be separated into six even ($g$) and six odd ($u$) irreps. For simplicity, here we display the character table for $D_6$, which is the same as the character table for $D_{6h}$ up to the introduction of symmetry operations composed with inversion and the identification of the irreps as even $(g)$ or odd $(u)$. The identification of the characters or figures with basis functions in terms of $(x,y,z)$ coordinates can be made in the same way as discussed for the case of $D_{4h}$ symmetry above.

\begin{table}[h]
\begin{center}
    \begin{tabular}{c}
\includegraphics[scale=0.4, keepaspectratio]{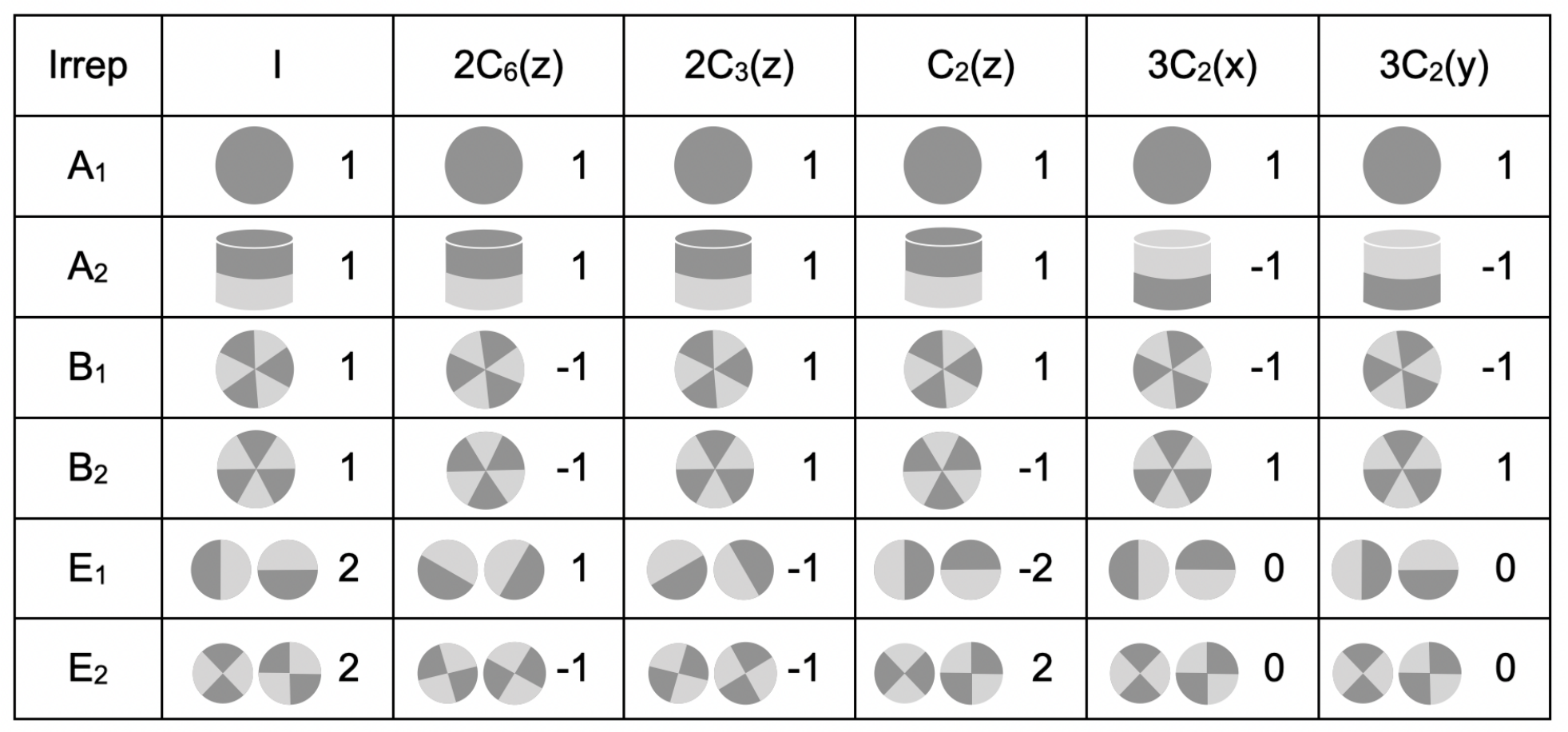}
       \end{tabular}
        \end{center}
\caption{\label{Tab:CTD6} 
Illustrated character table of the group $D_6$ with images accompanying the numerical entries. The dark/bright gray colors indicate regions with a +1/-1 value. The circles are in the $xy$-plane and the cylinder axis is along the $z$-axis.}
\end{table}

Concerning the groups $D_{3h}$ and $D_{3d}$, we can think of them as subgroups of $D_{6h}$ in order to derive the corresponding irreducible representations. The group $D_{3h}$ can be obtained by eliminating the operations $2C_6(z)$, $C_2(z)$, $3C_2(y)$, $i$, $2S_6$, and $3\sigma_d$, leaving us with twelve operations organized in six conjugacy classes, what leads to six irreducible representations. The elimination of the columns of the character table of $D_{6h}$ associated with the mentioned operations leads to the following mapping of the irreps from $D_{6h}$ to $D_{3h}$: $\{A_{1g}, B_{1u} \}\rightarrow A'_1$, $\{A_{2g}, B_{2u} \}\rightarrow A'_2$, $\{B_{1g}, A_{1u} \}\rightarrow A''_1$, $\{B_{2g}, A_{2u} \} \rightarrow A''_2$, $\{E_{1g}, E_{2u} \}\rightarrow E''$, and $\{E_{2g}, E_{1u}\} \rightarrow E'$. Conversely, the group $D_{3d}$ can be obtained by eliminating the operations $2C_6(z)$, $C_2(z)$, $3C_2(y)$, $2S_3$, $\sigma_h$ and $3\sigma_v$, leaving us again with twelve operations organized in six conjugacy classes and six irreducible representations. The elimination of the columns of the character table of $D_{6h}$ associated with the mentioned operations leads to the following mapping of the irreps from $D_{6h}$ to $D_{3d}$: $\{A_{1g/u}, B_{1g/u} \} \rightarrow A_{1g/u}$, $\{A_{2g/u}, B_{2g/u}\} \rightarrow A_{2g/u}$, $\{E_{1g/u}, E_{2g/u}\} \rightarrow E_g/u$ in the even/odd sector. 

The mappings above indicate that all the groups mentioned in Table \ref{Tab:Materials} are associated with two-dimensional irreps. The superconducting states developing in these materials can therefore be chiral, as we discuss in more detail in the next section. Note that the  point group $D_{2h}$, associated with UTe$_2$, can be obtained from $D_{4h}$ by preserving only $C_2(z)$, $C_2(z)$ and $i$, what leads to the following mapping of irreps: $\{A_{1g/u}, B_{1g/u} \} \rightarrow A_{1g/u}$, $\{A_{2g/u}, B_{2g/u} \} \rightarrow B_{1g/u}$, and $E_{g/u} \rightarrow B_{2g/u} + B_{3g/u}$. The last correspondence indicates that there are only one-dimensional irreducible representations in $D_{2h}$. This is expected, as this group encodes only two-fold rotations and inversion symmetries, which do not mix coordinates. In this context, we would not expect chiral superconductivity to develop in UTe$_2$. We hope this point will become clear in the next sections.

\section{Unconventional superconductivity from a group theory perspective}\label{Sec:Unconventional}

We now discuss the notion of unconventional superconductivity from the group theory perspective. Every superconducting order parameter should be associated with a given irreducible representation of the point group of the material of interest. As Cooper pairing is a phenomenon that occurs in momentum space, we can also think of these operations in momentum coordinates $(k_x, k_y, k_z)$. Order parameters transforming trivially under all point symmetry operations and time-reversal symmetry are considered conventional. The simplest example consists of an order parameter that is isotropic in space, which in the cases discussed above would be associated with the irrep labeled by $A_{1g}$ for $D_{4h}$, $D_{6h}$, and $D_{3d}$, and with the irrep labeled by $A'_1$ in $D_{3h}$. Experimental signatures of conventional superconducting states (as initially proposed by the BCS theory) are an activated behavior of the temperature dependence of the specific heat and robustness against nonmagnetic disorder.

Order parameters associated with a nontrivial irrep are always said to be \emph{unconventional}, namely, these transform nontrivially under some of the point group operations. This generally implies the presence of line nodes in the superconducting gap, namely, values of momenta for which the order parameter must go to zero as it must change sign. For example, an unconventional order parameter in a material with $D_{4}$ point group symmetry can be associated with the irrep $B_{1g}$: $\Psi_{B_{1}}(\bold{k}) = \Psi_0 (k_x^2-k_y^2)$, usually referred to as a \emph{d-wave} superconducting order parameter (here $\Psi_0$ is a constant associated with the magnitude of the order parameter). From the character table, we know that an order parameter in this irrep must change sign under $C_4(z)$ and $C_2(d)$, which implies that it must go to zero along the diagonals $k_x=\pm k_y$. Examples of the experimental signatures of symmetry-imposed nodes in the order parameter are the power-law dependence of the specific heat on temperature and the fragility of the superconducting state in the presence of nonmagnetic impurities \cite{Sigrist1991}. 

Unconventional superconducting states associated with two-dimensional irreps can display even more unusual experimental signatures \cite{Sigrist1991}. The freedom to select the relative magnitude and phase of the two components of the order parameter leads to fundamentally distinct types of order parameters in the same symmetry channel:

\begin{itemize}
\item \emph{Nematic:} If the order parameter is nematic, it develops as a linear combination of the two basis functions in a way that its amplitude breaks some of the symmetries of the normal state. Nematic order parameters can be favored in the vicinity of structural distortions, such that a lattice distortion can lead to further energy gain as the system goes into the superconducting state \cite{Sigrist1991}. One example of such a superconducting state in the context of $D_{4}$ point group symmetry is $\Psi_{E(N)}(\bold{k}) =  \Psi_0 k_x$, where $\Psi_0$ is a constant;

\item \emph{Chiral:} If the order parameter is chiral, it develops as a complex superposition of the two basis functions, which breaks time-reversal symmetry. One example of such a superconducting state in the context of $D_{4}$ point group symmetry is $\Psi_{E(C)}(\bold{k}) =  \Psi_0 (k_x+ i k_y)$, usually referred to as the \emph{chiral p-wave} state (here we assume no SOC and refrained to discuss the spin configuration in the Cooper pair, which must be in the triplet sector \cite{Sigrist1991}). Another interesting way to write this order parameter is $\Psi_{E(C)}(\bold{k}) =  \Psi_0 |\bold{k}|e^{i\theta_{\bold{k}}}$, where $\theta_{\bold{k}}$ parametrizes the angle associated with the momentum around the Fermi surface. In this case we can associate the order parameter with one unit of angular momentum, as it acquires a phase of $2\pi$ as it circles the Fermi surface (assumed to be circular in two dimensions). In the chiral case, the amplitude of the order parameter, $|\Psi_{E(C)}(\bold{k})| = \Psi_0 \sqrt{k_x^2+k_y^2}$, does not break the symmetries of the normal state.  The development of a chiral order parameter is usually favored by the fact that a complex superposition of two components releases the nodes in the superconducting gap and increases the condensation energy (the energy the system gains for going into the superconducting state). The fewer nodes in the gap, the more electrons pair, and more energy is gained once the system goes to the superconducting state \cite{Kallin2016}.
\end{itemize}

\begin{table}[h]
\begin{center}
    \begin{tabular}{c}
   \includegraphics[width=0.95\linewidth, keepaspectratio]{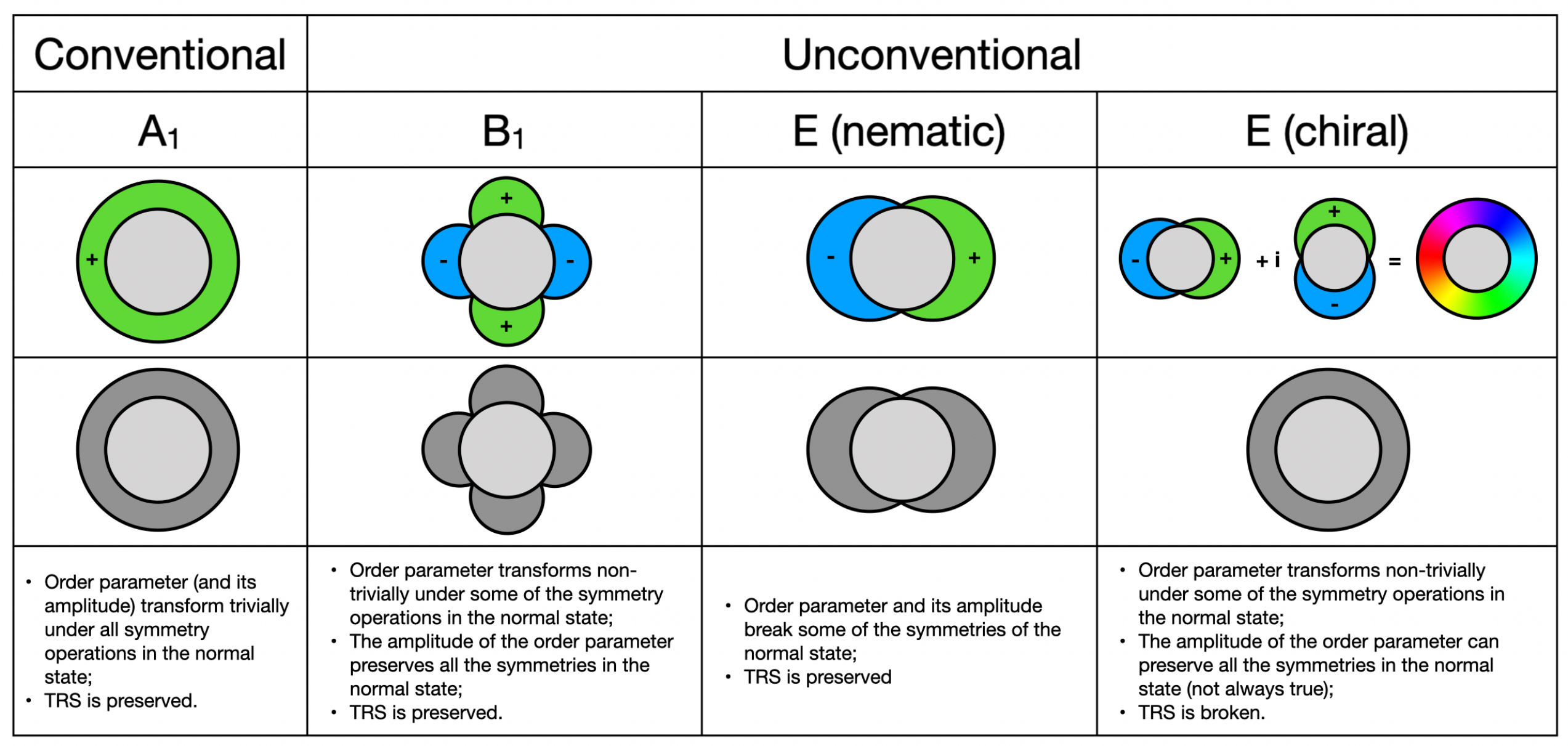}
       \end{tabular}
        \end{center}
\caption{\label{Tab:Unconventional} Summary of properties of conventional and unconventional superconductors from the perspective of the irreps of the point group $D_4$.}
\end{table}

\section{Landau Theory of Phase Transitions}\label{Sec:Landau}

With the group theory background presented above, we can now discuss the Landau theory of phase transitions, focusing on superconductivity developing in systems with a given point group symmetry.

The Landau theory of phase transitions assumes that the transition from a disordered to an ordered phase of matter can be captured by a free energy written as a Taylor expansion on the respective order parameter. The formalism only assumes that the free energy is analytic in the order parameter and that it should obey all the symmetries of the Hamiltonian in the disordered state, such that the symmetry breaking occurs spontaneously. Importantly, the development of the ordered phase is associated with some symmetry breaking captured by the order parameter.

In the context of superconductivity, the most straightforward Landau theory that describes the superconducting phase transition is written in terms of the order parameter as a single complex number $\Psi$. The motivation for this choice came from the notion of a Bose-Einstein condensate, in which all particles can occupy the same single-particle quantum state. Consequently, the free energy should not depend on the global phase of the order parameter as global phases of quantum states are not observable. The most general free energy, in this case, can be written in terms of powers of the amplitude of the order parameter:
\begin{eqnarray}
F = F_0(T)+\alpha(T) |\Psi|^2 + \frac{\beta}{2}|\Psi|^4+ ...
\end{eqnarray}
where $F_0(T)$ is the free energy of the normal state, the coefficients $\alpha$ and $\beta$ are introduced phenomenologically, and the dots correspond to higher order terms in $\Psi$. In particular, $\beta$ should be positive for the stability of the ordered phase, meaning the free energy is bound from below. We can take explicitly $\alpha(T) = a (T/T_c-1)$, with $a$ a positive phenomenological constant, $T$ the temperature, and $T_c$ the superconducting critical temperature. Note that the factor $\alpha$ is positive for $T>T_c$ and negative for $T<T_c$. Minimizing the free energy with respect to $\Psi$ for $T>T_c$ gives us a single minimum for $\Psi = 0$, while minimizing the free energy for $T<T_c$ gives us a ring of minima with order parameter amplitude $|\Psi| = \sqrt{-\alpha(T)/\beta}$ and arbitrary phase.

This form for the free energy functional is appropriate for the description of single-component order parameters. As we have discussed above, there are order parameter realizations with multiple components associated by symmetry. In these cases, the free energy needs to be generalized, including new types of fourth-order terms which depend on the point group symmetry. Before discussing cases with two components, we briefly discuss the nature of the fourth-order term in the simplest case from a group theory perspective. If the order parameter is single-component, it is associated with a one-dimensional irreducible representation. Note that the entries in the character table associated with one-dimensional irreps are always $\pm 1$. With these, we can already infer how the product of order parameters transforms. Suppose we take $|\Psi|^2$, the product of the order parameter with its complex conjugate. In that case, we can infer how the product transforms by taking the product of the characters associated with the irrep of $\Psi$. As an example, we can take the $B_1$ irrep in the point group $D_4$, as shown on the fourth line of Table \ref{Tab:CTD4}. If we take the product of the entries in each column with their complex conjugate, we always find $+1$. This means that $|\Psi|^2$ transforms according to the trivial irreducible representation, $A_1$, as all its characters are equal to $+1$. More formally we could write $B_1 \otimes B_1 = A_1$. This means that $|\Psi|^2$ can be written in the free energy, as it is invariant under all transformations of the point group. It follows that $|\Psi|^4$ is also symmetry allowed.

Moving now to the case of two components associated with a two-dimensional irreducible representation, we can do a similar exercise to find out what terms are symmetry allowed in the free energy functional. Let us take as an example the irrep $E$ in $D_4$, as shown in the sixth line of Table \ref{Tab:CTD4}. In this case, one of the components can be chosen as transforming as the coordinate x, which we label as $\Psi_x$, and another as the coordinate $y$, labeled as $\Psi_y$. This is conveyed by the associated images in Table \ref{Tab:CTD4}. We need a real product of the two components that transforms trivially for the second-order term. The only possibility is to take the square of the absolute value of the associated vector $\vec{\Psi} = (\Psi_x,\Psi_y)$, as $|\vec{\Psi}|^2 = (\Psi_x,  \Psi_y).(\Psi_x^*, \Psi_y^*) = |\Psi_x|^2 +  |\Psi_y|^2$. We can briefly explicitly check that these are all the possible combinations of $\Psi_x$ and $\Psi_y$ that transform trivially under all point group operations. Under $C_4(z)$ rotations we have $\Psi_x \rightarrow \Psi_y$ and $\Psi_y \rightarrow - \Psi_x$, under $C_2(z)$ rotations we have $\Psi_x \rightarrow -\Psi_x$ and $\Psi_y \rightarrow - \Psi_y$, under $C_2(x)$ rotations we have $\Psi_x \rightarrow \Psi_x$ and $\Psi_y \rightarrow - \Psi_y$, and under $C_2(d)$ rotations we have $\Psi_x \rightarrow \Psi_y$ and $\Psi_y \rightarrow \Psi_x$. All these correspondences keep the $|\vec{\Psi}|^2$ invariant, so this is an allowed term in the free energy. Conversely, the second order term $\Psi_x \Psi_y^* - \Psi_x^* \Psi_y$ pick up minus signs under $C_4(z)$, $C_2(x)$, and $C_2(d)$ rotations, so it does not transform trivially under some of the point group operations and therefore is not allowed in the free energy.

Concerning the fourth order terms, we already know that $|\vec{\Psi}|^4$ is symmetry allowed, as this is simply the square of the second order term. Expanding this term we find $|\vec{\Psi}|^4 = |\Psi_x|^4 + 2|\Psi_x|^2 |\Psi_y|^2 + |\Psi_y|^4$. Note that the term $|\Psi_x|^2 |\Psi_y|^2$ also transforms trivially under all symmetry operations listed above. Therefore this term is allowed to be present in the free energy, independently of $|\vec{\Psi}|^4$. This means we have at least two qualitatively distinct terms in the free energy that should come with different phenomenological parameters. Are these all the symmetry-allowed terms? There is one more possibility for the $E$ irrep in the $D_4$ point group. We can guess its form by taking the square of the second order term that was not allowed in the free energy, as it gained a minus sign under some symmetry transformations. Note that the square of this term is invariant under all symmetry transformations, as $(-1)^2=+1$. Expanding this term we find $(\Psi_x \Psi_y^* - \Psi_x^* \Psi_y)^2 = (\Psi_x)^2 (\Psi_y^*)^2 - 2 |\Psi_x|^2 |\Psi_y|^2 + (\Psi_x^*)^2 (\Psi_y)^2$. As we know that $|\Psi_x|^2 |\Psi_y|^2$ itself is invariant, we conclude that $(\Psi_x \Psi_y^*)^2 + (\Psi_x^*\Psi_y)^2$ is also invariant independently. This means that we could have taken either $(\Psi_x \Psi_y^* \pm \Psi_x^* \Psi_y)^2$. In the literature, the choice with a minus has been used more extensively, in particular in \cite{Sigrist1991}, so we stick to the minus sign choice to write the most general form of the free energy for a two-component superconducting order parameter in a system with $D_4$ symmetry:
\begin{eqnarray}
F_{D_{4h}} = F_0(T)+\alpha(T) |\vec{\Psi}|^2 + \beta_1 |\vec{\Psi}|^4+ \beta_2 (\Psi_x \Psi_y^* - \Psi_x^* \Psi_y)^2 +\beta_3|\Psi_x|^2 |\Psi_y|^2 + ....,
\end{eqnarray}
where $\beta_i$ with $i=1,2,3$ are phenomenological parameters. Note that as $D_{4h}$ is an extension of $D_4$ by including inversion symmetry, the form of the free energy for the $D_{4h}$ point group is the same, as all powers of the free energy are even in the order parameter, so this functional form is valid for both $E_g$ and $E_u$ irreps in $D_{4h}$.

A similar discussion would allow us to write down the most general form for the free energy for the four different two-dimensional irreps ($E_{1g}$, $E_{2g}$, $E_{1u}$, and $E_{2u}$) in $D_{6h}$ (also valid for $D_{3h}$ and $D_{3d}$):
\begin{eqnarray}
F_{D_{6h}} = F_0(T)+\alpha(T) |\vec{\Psi}|^2 + \beta_1 |\vec{\Psi}|^4+ \beta_2 (\Psi_x \Psi_y^* - \Psi_x^* \Psi_y)^2 + ....
\end{eqnarray}

The coefficients $a$ and $\beta_i$ for each case can, in principle, be evaluated by a microscopic theory. Here we take a pragmatic path and discuss the conditions on these parameters for the development of a chiral superconducting state. For both the $D_{4h}$ and $D_{6h}$ scenarios, we see that the term proportional to $\beta_2$ might be the key to the development of TRSB in the superconducting state, as it is the only term that depends explicitly on the relative phase of the two components. We want to determine the accompanying factor in terms of the order parameter components to minimize the free energy. If we parametrize the two components in terms of one overall amplitude ($\Psi_0$), one angle $(\theta)$ controlling the relative amplitude and another angle $\phi$ corresponding to the relative phase of the two components: $\vec{\Psi} = (\Psi_1,\Psi_2) = \Psi_0(\cos\theta, \sin\theta e^{i\phi})$, we find that the $\beta_2$ term is equal to $-\beta_2 |\Psi_0|^4 \sin^2(2\theta)\sin^2\phi$. If $\beta_2>0$, this term in minimized for $\phi = \pm \pi/2$, while if $\beta_2<0$, this term is minimized by $\phi = \{0, \pm \pi\}$. A necessary condition for developing chiral superconductivity is a positive $\beta_2$ coefficient.

The condition on $\beta_2$  for a complex relative phase between the two parameters needs to be supplemented by the condition on other terms that guarantee that the two components develop a finite amplitude. This can be seen from the $\beta_3$ term in $D_{4h}$ but would require sixth order terms for the case of $D_{6h}$ symmetry. Focusing on the $D_{4h}$ scenario, using the parametrization suggested above, the terms with $\beta_2$ and $\beta_3$ combined lead to $|\Psi_0|^4\sin^2(2\theta)[\beta_3/4 - \beta_2 \sin^2\phi]$. A positive factor within the squared brackets implies that the free energy is minimized for $\theta = \{0, \pm \pi/2\}$, while a negative factor implies minimization of the free energy with $\theta = \pm \pi/4$. Therefore a second condition for the development of chiral superconductivity (at least in the context on the $D_{4h}$ point group) has been identified: $[\beta_3/4 - \beta_2 \sin^2\phi]<0$.

Within the Landau theory of phase transitions, we can also consider how the superconducting order parameter couples to lattice deformations and what are the experimental signatures that are uniquely associated with multi-component order parameters stemming from multi-dimensional irreps. Lattice deformations can have two characters: compressive strains, for which the lattice parameters are changed, but no lattice symmetry is broken, and shear strains, which break some lattice symmetries. Using the group theory notions introduced in this work, assuming point group $D_{4h}$, we can say that compressive strains are associated with the trivial $A_{1g}$ irrep, while the shear strains are associated with the nontrivial irreps. We can then write new terms in the free energy functional, coupling the order parameters to lattice distortions, following  the requirement that their product transforms trivially under all point group symmetries. Note that linear couplings of lattice deformations to the order parameter are still forbidden, as the order parameter products should always appear in a real combination. We can then consider couplings that are linear in strain and quadratic in the order parameter. We have learned above that if the order parameter belongs to a one-dimensional irrep, the only allowed product in second order is $|\Psi|^2$, which belongs to the trivial irrep. This means that order parameters associated with a one-dimensional irrep can only couple to compressive strain. In contrast, order parameters with two components can be combined in second-order products that do not necessarily transform as the trivial representation. For example, we can consider the combination $|\Psi_x|^2 - |\Psi_y|^2$, which transforms as the $B_{1g}$ irreducible representation in $D_{4h}$. This term alone cannot appear in the free energy functional, but the product $\epsilon_{B1g}(|\Psi_x|^2 - |\Psi_y|^2)$ can. Another example would be the coupling $\epsilon_{B2g}(\Psi_x \Psi_y^* + \Psi_x^* \Psi_y)$. Here $\epsilon_{B1g} = \epsilon_{xx}-\epsilon_{yy}$ and $\epsilon_{B2g} = 2\epsilon_{xy}$ correspond to strain with $B_{1g}$ and $B_{2g}$ symmetry, respectively, where $\epsilon_{ij}$ are strain components, characterizing the relative change in lengths and angles of the lattice under the specific deformation.

These new terms in the free energy lead to two important and experimentally verifiable consequences. The first consequence  concerns the propagation of sound in the material. Only multi-component superconductors can manifest a discontinuity in the velocity of sound waves associated with nontrivial irreps at the superconducting transition temperature. The second is the splitting of the superconducting transition under certain types of uniaxial strain. If the strain breaks the symmetry between the two components, their superconducting critical temperature is not the same anymore. One example consists of $B_{1g}$ strain, corresponding to uniaxial compression along the $x$-direction in $D_{4h}$, such that the symmetry between the $x$ and $y$ directions is explicitly broken. In absence of strain the $x$ and $y$ components are degenerate (they have the same superconducting critical temperature) by symmetry, possibly leading to a chiral superconducting order parameter. Under strain, the critical temperature for the $x$ component is enhanced, while for the $y$ component it is reduced (or vice-versa), and a chiral state cannot be established right below the critical temperature, as it requires that both components develop a finite value. The recent experiments on ultrasound attenuation \cite{Ghosh2021} and $\mu$SR under strain in Sr$_2$RuO$_4$ \cite{Grinenko2021b, Grinenko2021a} beautifully reveal this expected phenomenology for a multi-component order parameter.

\section{Conclusion and discussion}

In this article, we have highlighted recently proposed materials to host chiral superconductivity. We focused on systems with direct evidence of time-reversal symmetry breaking in the superconducting state. However, multiple other materials and heterostructures have been recently theoretically proposed to host this fascinating state of matter, posing a welcoming challenge to experimental colleagues working on the synthesis and characterization of materials at low temperatures.

After introducing basic notions of group theory, we briefly accounted for the notion of unconventional superconductivity from a lattice symmetry perspective. We discussed how the multi-component nature of chiral superconducting order parameters is intimately associated with two-dimensional irreducible representations. 

 In the last section, we took a pragmatic approach. We discussed the stability of chiral superconductivity based on generalized Landau functionals for multi-component order parameters for two distinct point groups. The study of the details of the pairing mechanism and the evaluation of the phenomenological Landau parameters from microscopic theories are desirable and promising directions of work in the following years, given the new plethora of uncovered time-reversal symmetry-breaking superconductors in recent times.

\section*{Funding and Acknoledgements}
The author is supported by the Swiss National Science Foundation through the Ambizione Grant No. 186043. The author thank Manfred Sigrist for discussions.

\bibliographystyle{unsrt}
\bibliography{IntroChiralSCs}{}

\end{document}